\documentclass{optica-article}

\journal{opticajournal} 

\articletype{Research Article}

\usepackage{lineno}

\begin{document}

\title{Influence of mechanical resonances on the linearity of adiabatic frequency conversion in whispering gallery resonators}

\author{Alexander Mrokon,\authormark{1,*} Till Wachweger,\authormark{1} Dongsung Shin,\authormark{1} Karsten Buse,\authormark{1,2} and Ingo Breunig\authormark{1,2}}

\address{\authormark{1}Laboratory for Optical Systems, Department of Microsystems Engineering - IMTEK, University of Freiburg, Georges-Köhler-Allee 102, Freiburg, 79110, Germany\\
\authormark{2}Fraunhofer Institute for Physical Measurement Techniques IPM, Georges-Köhler-Allee 301, Freiburg, 79110,Germany\\}

\email{\authormark{*}alexander.mrokon@imtek.uni-freiburg.de} 


\begin{abstract*} 
Adiabatic frequency conversion enables fast and efficient tuning of laser light by coupling it into an optical resonator whose eigenfrequency is varied on a timescale shorter than its photon lifetime. In this regime, the optical frequency follows the cavity resonance, allowing frequency shifts of several hundred gigahertz within sub-microsecond time - independent of optical power and without phase-matching constraints. While a linear dependence of the cavity resonance on a control parameter (e.g., applied voltage) suggests that arbitrary temporal signals could be linearly transferred to optical frequency changes, we show that this assumption fails near mechanical resonances of the resonator. Using a millimeter-sized lithium niobate whispering gallery resonator with a pronounced mechanical mode at 10.5~MHz, we observe strong deviations from linearity even when higher harmonics of the control signal coincide with this resonance. The experimental results are in excellent agreement with theoretical predictions. They demonstrate that mechanical resonances impose intrinsic limits on the linearity of adiabatic frequency conversion and other frequency control schemes based on the variation of the eigenfrequency of an optical cavity. 
\end{abstract*}

\section{Introduction}
Adiabatic frequency conversion (AFC) is a powerful approach for rapidly tuning the frequency of laser light. In this scheme, light is first coupled into an optical resonator whose eigenfrequency is subsequently varied on a timescale shorter than the cavity decay time. As a result, the optical field stored within the resonator experiences a frequency modulation that follows the temporal variation of the cavity eigenfrequency \cite{SatoshiWavelength}. Notably, this process operates efficiently, regardless of the optical input power, and does not require any phase-matching conditions to be fulfilled.

In the first experimental demonstration, laser light at a wavelength of 1.5~\textmu m was coupled into a silicon ring resonator. A short laser pulse generated mobile charge carriers in the silicon, thereby changing its refractive index and inducing frequency shifts of up to 300~GHz within a picosecond, corresponding to a tuning rate of $3\times10^5$~GHz/ns \cite{LipsonAFC}. This remarkable rate illustrates the potential of AFC for ultrafast frequency control. Subsequent implementations have demonstrated AFC in photonic crystal cavities \cite{TanabeElectroOptic, TanabeDynamicRelease, KondoAdiabatic}, integrated waveguides \cite{KampfrathUltrafast, UphhamOnTheFly, KondoDynamicWavelength}, and fiber Bragg grating resonators \cite{KabakovaSwitching}. Its applicability to the quantum regime has also been confirmed through frequency conversion of single photons \cite{PrebleSinglePhotonAFC, HongTangAFCPhoton}.

Electro-optically driven adiabatic frequency conversion based on the Pockels effect is particularly attractive, as it enables tuning the eigenfrequency of an optical resonator through a readily accessible control parameter, an applied voltage. The induced frequency shift scales proportionally with the driving voltage, allowing both up- and down-shifts of the optical frequency with high precision \cite{YannickAFC,CardenasPICAFC}. Since the underlying effect is strictly linear, one might state that any temporally varying voltage signal is transferred linearly into a corresponding frequency shift of the optical field. Experimentally, this assumption has so far been supported by the successful conversion of linearly ramped voltage signals into frequency chirps \cite{MrokonAFCFMCW}. The high transfer fidelity even enabled applications in frequency-modulated continuous-wave LiDAR. Hence, the abovementioned statement appears, at first glance, to be well justified.

However, this assumption holds only if no other effects come into play. It should be recalled that adiabatic frequency conversion involves repeatedly filling the resonator with light, shifting its eigenfrequency within the photon lifetime, and refilling it again. Consequently, the driving voltage signal becomes periodic on a timescale comparable to the photon lifetime. Typical photon lifetimes in whispering gallery resonators made of bulk crystals range from 10 to 100 ns, while chip-integrated cavities typically exhibit values between 0.1 and 10 ns. Accordingly, the associated voltage signals contain equidistant spectral components separated by 10 - 100 MHz for bulk resonators and 0.1 - 10~GHz for on-chip implementations.

The electro-optic response is known to vary significantly near mechanical resonances of the resonator material \cite{Abarkan03,Takeda12}. For bulk whispering-gallery resonators, these resonances typically occur between 1 and 100~MHz \cite{WenleWengPiezo}, whereas on-chip devices exhibit mechanical modes in the 0.1~MHz - 10~GHz range \cite{HongTangAlNChip,HongTangElectromech,HongTangSubTHz}. Thus, it is by no means unlikely that spectral components of the applied voltage signal coincide with these resonance frequencies. Under such conditions, the resonator’s response to an applied voltage can no longer be considered constant, and a strictly linear transfer from voltage to optical frequency shift is not guaranteed.

In this work, we systematically investigate how mechanical resonances influence the linearity of electro-optically driven adiabatic frequency conversion. As a model system, we employ a millimeter-scale lithium niobate whispering gallery resonator exhibiting a pronounced mechanical resonance near 10 MHz. By applying temporally periodic voltage signals with well-defined spectral components, we probe the interaction between the electrical drive and the mechanical response of the resonator up to the ninth harmonic of the driving frequency. Furthermore, we identify a strategy to maintain linear frequency control even in the presence of mechanical resonances.

\section{Theoretical considerations}

\begin{figure*}[htpb]
\centering
\includegraphics[width = \textwidth]{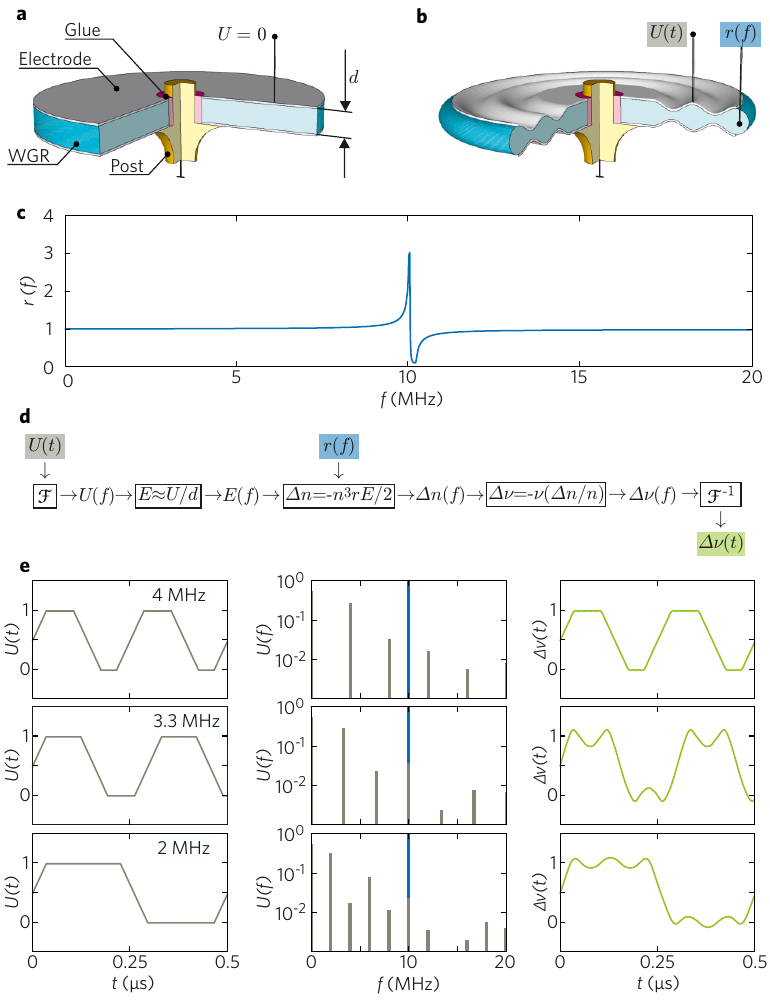}
\caption{\textbf{a} Schematic illustration of the lithium niobate whispering gallery resonator (WGR) mounted on a post, without applied voltage. \textbf{b} The applied voltage $U(t)$ induces mechanical deformation via the piezoelectric effect. \textbf{c} Frequency response of the WGR to an applied voltage $r(f)$ showing a mechanical eigenmode near 10~MHz. \textbf{d} Signal flow model illustrating the effect of the voltage waveform $U(t)$ on the optical frequency shift $\Delta\nu(t)$ via the frequency-dependent electro-optic response. \textbf{e} Example voltage waveforms (left), their spectral components (center), and the resulting optical frequency shifts (right), showing the influence of resonance overlap on modulation linearity.}
\label{figTheorie}
\end{figure*}

A simplified scheme is illustrated in Figure \ref{figTheorie}a and \ref{figTheorie}b. We consider a disk-shaped whispering gallery resonator (WGR) made of a non-centrosymmetric crystal mounted on a metal post. A temporally varying voltage signal $U(t)$ is applied between the top and bottom electrodes, which are separated by the distance $d$. When $U(t)\neq0$, the crystal deforms through the inverse piezoelectric effect, and the resulting strain modifies the refractive index via the elasto-optic effect \cite{Newnham}. This coupling leads to strong variations in the electro-optic response when the applied voltage signal excites a mechanical resonance of the crystal, as sketched in Fig. \ref{figTheorie}c. For simplicity, we consider a single mechanical mode that enhances the electro-optic coefficient $r$ by a factor of three near the frequency $f=10$~MHz.

It should be noted that at low frequencies, the electro-optic response consists of two contributions: 
the so-called "true" (primary) and "false" (secondary) electro-optic coefficients.

The primary contribution $r_{\text{true}}$ arises directly from the Pockels effect, that is, the linear change of the refractive index induced by an applied electric field.
On the other hand, the secondary contribution $r_{\text{false}}$, appears indirectly. 
The electric field first causes a deformation of the crystal due to the piezoelectric effect, which, through the elasto-optic effect, leads to an additional change in the refractive index. 
This contribution can be expressed as $r_{(\text{false})} = pd(f)$ where $p$ are the elasto-optic coefficients 
and $d(f)$ the frequency-dependent piezoelectric coefficients of the material.
The apparently measured electro-optic coefficient is thus given by $r = r_{\text{true}} + r_{\text{false}}.$

The effects of the false electro-optic coefficients generally diminish and become negligible at higher frequencies \cite{Newnham}. In lithium niobate, the influence of the piezoelectric contribution is one order of magnitude smaller than the true electro-optic effect \cite{MinetElectricField}.
Nevertheless, it is important to be aware of its presence as it alters the geometry and thus changes the resonance frequency of the WGR.
In our setup, all contributions are included in the frequency-dependent electro-optic response $r(f)$, as we investigate the reaction of the WGR to an applied electric field and can only observe the resulting shift in the resonance frequency.

Having defined the driving voltage signal $U(t)$ and the frequency-dependent electro-optic response 
$r(f)$, we now determine the resulting frequency shift $\Delta\nu(t)$ in electro-optically driven adiabatic frequency conversion. The corresponding signal chain is illustrated in Fig.~\ref{figTheorie}(d). First, the voltage signal is Fourier-transformed to obtain its spectral representation $U(f)$. Dividing by the electrode spacing $d$ yields the electric field $E(f)$ inside the WGR. For simplicity, the disk-shaped resonator is approximated as a parallel-plate capacitor, neglecting the slight field reduction caused by the curvature of the rim \cite{MinetElectricField}.

The electric field $E(f)$ modifies the refractive index $n$ of the material via the linear electro-optic (Pockels) effect according to
\begin{equation}
\Delta n(f) = -\frac{1}{2}n^3r(f)E(f)\;.
\end{equation}

This refractive-index change alters the optical eigenfrequency of the resonator and, consequently, the frequency $\nu$ of the light stored inside the cavity. The corresponding frequency shift in the spectral domain is given by $\Delta\nu(f) = -(\nu/n)\Delta n(f)$. Applying the inverse Fourier transform finally yields the temporal evolution $\Delta\nu(t)$ of the optical frequency shift.

We now apply the signal chain to three representative voltage waveforms, as illustrated in Fig.~\ref{figTheorie}e. First, we consider a trapezoidal voltage signal $U(t)$ with a repetition rate of 4 MHz. Its spectrum contains components at $f=0,4,8,12,\ldots$~MHz, i.e., none coincide with the mechanical resonance at 10~MHz. Consequently, the electro-optic response $r(f)$ can be regarded as constant, and the resulting frequency change $\Delta\nu(t)$ retains a trapezoidal shape that faithfully reproduces the driving voltage $U(t)$ via \cite{YannickAFC}
\begin{equation}
\Delta\nu(t) = \frac{1}{2}\nu n^2 r \frac{U(t)}{d}\;. \label{eq:Linear}
\end{equation}

Next, we reduce the repetition rate to 3.3~MHz, resulting in spectral components at $0, 3.3, 6.7, 10,\ldots$~MHz, such that the third harmonic coincides with the mechanical resonance. The corresponding component of the electro-optic response is enhanced by a factor of three compared to the others. As a result, the frequency shift deviates strongly from the trapezoidal waveform and exhibits an additional periodic modulation at 10~MHz in regions of constant voltage.

Finally, when the repetition rate is further decreased to 2~MHz, the fifth harmonic of the voltage spectrum overlaps with the mechanical resonance. Again, the frequency change departs from the shape of the driving signal and shows a superimposed 10 MHz modulation. However, the amplitude of this deformation is smaller than in the previous case, since the fifth-order spectral component has a lower amplitude than the third-order one, and therefore contributes less despite the resonance enhancement.

These examples clearly demonstrate that the linear transfer between the driving voltage and the optical frequency shift can be significantly distorted when harmonic components of the voltage spectrum coincide with mechanical resonances of the resonator. Even if the driving signal itself is purely linear and periodic, the frequency-dependent electro-optic response introduces additional modulations in the resulting frequency shift. The strength of this deformation increases with the amplitude of the spectral components that overlap with the mechanical resonance. Hence, mechanical resonances represent a fundamental limitation for achieving perfectly linear electro-optic frequency control.

\begin{figure}[ht]
\centering
\includegraphics[width=\linewidth]{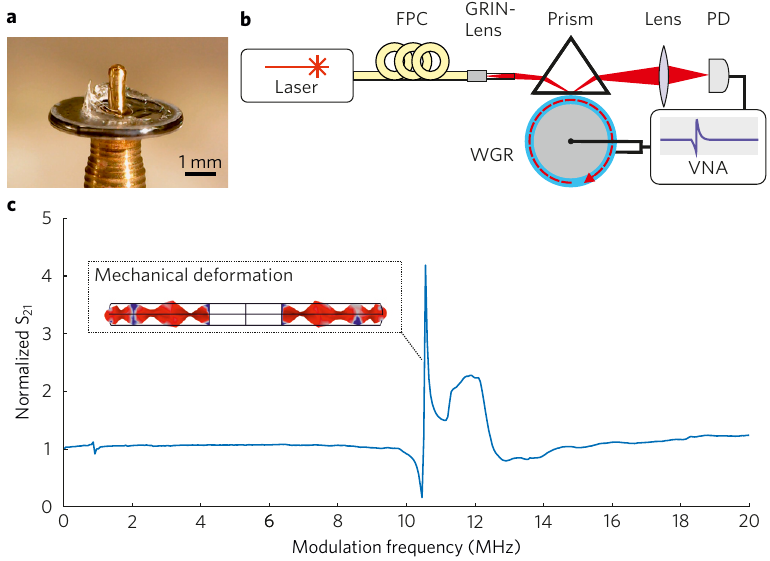}
\caption{ \textbf{a} Photograph of the lithium niobate resonator mounted on a brass post. \textbf{b} Experimental setup for characterizing the piezoelectric response of the WGR. A laser emitting at 1560~nm is coupled into the resonator using a fiber polarization controller (FPC), a GRIN lens and a rutile prism. A vector network analyzer (VNA) applies a voltage to the electrodes. A photodiode (PD) detects the beat signal, which is recorded by the VNA. \textbf{c} Frequency dependent normalized electro-optic response $S_{21}$ of the WGR with a pronounced mechanical eigenmode near 10.5\,MHz. Inset: Simulated mechanical deformation pattern corresponding to this eigenmode.}
\label{figVNA}
\end{figure}


\section{WGR-Fabrication and characterization}

To experimentally verify the behavior described above, we fabricated a millimeter-scale WGR following the approach presented in Ref.~\cite{WernerGeometricTuning}. The device was made from a 0.3-mm-thick, single-crystalline z-cut lithium niobate wafer doped with 5~mol\% MgO. The top and bottom surfaces were coated with 150-nm-thick chromium electrodes.

The fabrication process starts with cutting a toroidal preform of 4~mm outer and 1~mm inner diameter using a femtosecond pulsed laser (388~nm wavelength, 200~fs pulse duration, 2~kHz repetition rate, 300~mW average power). The preform was subsequently mounted onto a brass post with a diameter of 1.5~mm. To minimize mechanical coupling between the WGR and the post, the contact area was reduced so that the resonator was largely surrounded by air, as illustrated in Fig.~\ref{figTheorie} a. The rim of the preform was then shaped using the same femtosecond laser in combination with a computer-controlled lathe, yielding a final major diameter of 3.8~mm and a minor diameter of 0.76~mm. Finally, the resonator rim was polished with a 50-nm diamond slurry to minimize optical scattering losses, increasing the intrinsic quality factor to $1.5~\times10^8$ (corresponding to approximately 1~MHz linewidth) for extraordinarily polarized light at 1560~nm wavelength, which is close to the absorption limit of lithium niobate \cite{LeidingerComparativeStudy}.

Once the WGR is fabricated, its mechanical resonances must be characterized, which are expected to occur in the 1 - 10~MHz range. The corresponding experimental setup is illustrated in Fig.~\ref{figVNA} b. Continuous-wave laser light at a wavelength of 1560~nm (linewidth $<$90~kHz) is coupled into the resonator via a rutile prism. The coupling gap is adjusted such that the loaded quality factor is on the order of $10^7$, corresponding to an optical linewidth in the tens of megahertz range. The resonator temperature is actively stabilized at 35$^\circ$C with residual fluctuations around one millikelvin. These measures ensure that the laser frequency can be stably blue-detuned from the optical resonance to the point of maximum slope in the transmitted signal detected by a photodiode.

Port~1 of a vector network analyzer (VNA) applies sinusoidal voltage signals with frequencies between 9~kHz and 20~MHz. The resulting temporal modulation of the resonator’s eigenfrequency produces a corresponding intensity variation at the photodiode output, which is connected to Port~2 of the VNA. 
Using this configuration, the measured normalized transmission parameter $S_{21}$ directly represents the normalized, frequency-dependent electro-optic response of the resonator, as shown in Fig.~\ref{figVNA} c. Between modulation frequencies of 10 and 12.5~MHz, the normalized response varies from 0.2 to 4.2, exhibiting a distinct resonance peak at 10.5~MHz. Finite-element simulations performed with COMSOL Multiphysics confirm that this feature corresponds to a piezoelectrically-induced mechanical mode, as illustrated in the inset of Fig.~\ref{figVNA} c. Outside the 10 - 12.5~MHz range, the electro-optic response remains nearly constant, varying by less than ±20~\%.

\section{Adiabatic frequency tuning}

\begin{figure}[htpb]
\centering
\includegraphics[width=\linewidth]{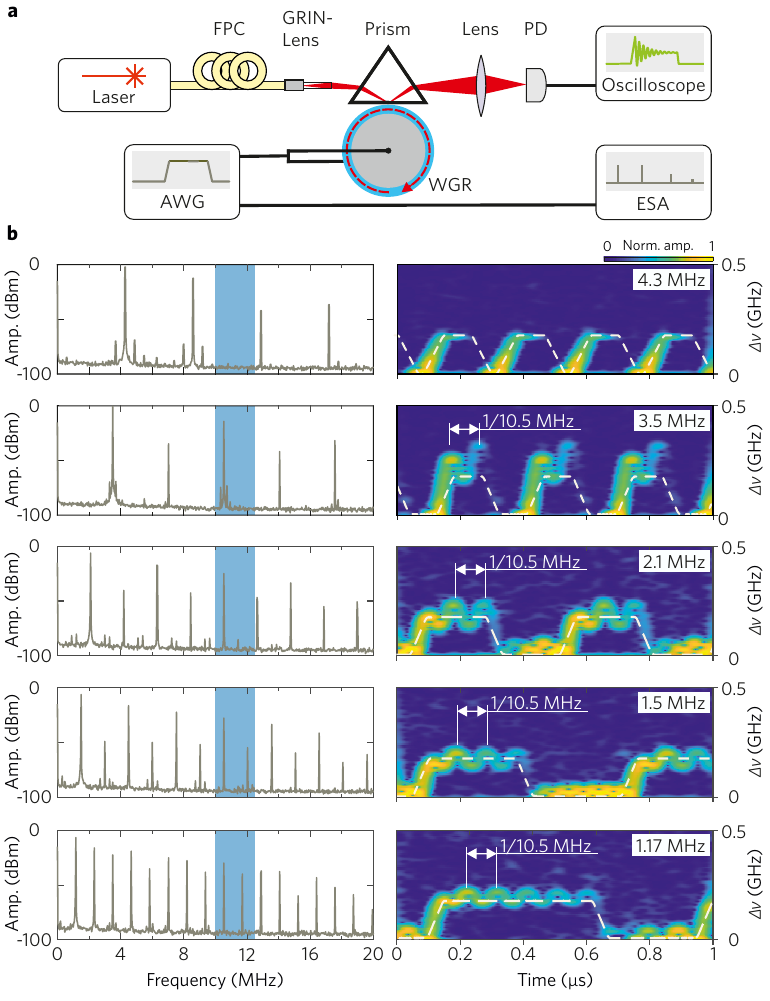}
\caption{\textbf{a} Experimental setup to investigate the influence of piezoelectric resonances on AFC in a lithium niobate WGR. The resonator is driven by an arbitrary waveform generator (AWG), and the signal is analyzed using an electrical spectrum analyzer (ESA). \textbf{b} Frequency spectra of the applied voltage signals (left) and the corresponding optical frequency shifts $\Delta\nu(t)$ (right) for five different drive frequencies: 4.3, 3.5, 2.1, 1.5, and 1.17~MHz. The blue-shaded region marks the mechanical resonance around 10.5~MHz. The white dashed lines indicate the expected behavior for linear adiabatic frequency conversion.}
\label{figAFC}
\end{figure}

In the second part of the experiment, we investigate how piezoelectrically driven mechanical resonances affect the linearity of adiabatic frequency conversion (AFC). The corresponding setup is shown in Fig.~\ref{figAFC}(a). Continuous-wave laser light at a wavelength of 1560~nm (optical frequency $\nu=192$~THz) is coupled into the WGR via a rutile prism. In contrast to the previous configuration, an arbitrary waveform generator (AWG) supplies a voltage signal $U(t)$ to the electrodes of the resonator and simultaneously to the input of an electrical spectrum analyzer (ESA), which records the spectral content $U(f)$ of the driving signal. The light, frequency-shifted to $\nu+\Delta\nu(t)$, exits the resonator and interferes with the unshifted laser light at $\nu$, producing a beat signal detected by a photodiode connected to an oscilloscope. From this beat signal, the instantaneous frequency shift $\Delta\nu(t)$ is extracted and compared with the temporal profile $U(t)$ of the driving voltage.

The AWG is configured to generate trapezoidal voltage waveforms: the voltage is initially set to zero to allow light coupling into the WGR, then ramped linearly from 0 to 5~V, held at 5~V, and linearly decreased back to zero. This waveform is periodically repeated with modulation frequencies of 4.3, 3.5, 2.1, 1.5, and 1.17~MHz. For a reliable and repeatable adiabatic frequency shift, it is essential that the WGR is effectively "empty" before being refilled with laser light at frequency $\nu$, i.e., no residual light remains at the previously shifted frequency $\nu+\delta\nu$. Therefore, the decay time for the light in the resonator must be adapted to the modulation period. To achieve this, the coupling gap between the prism and the WGR is adjusted such that the decay time at a modulation frequency of 4.3~MHz is approximately 200~ns. As the modulation frequency is reduced, the gap is gradually increased, thereby extending the decay time. At 1.17~MHz, the decay time reaches approximately 400~ns.

At 4.3~MHz, none of the frequency components of the voltage signal fall within the 10 - 12.5~MHz range, and thus the mechanical resonances are not expected to influence the frequency conversion process. In contrast, for modulation frequencies of 3.5, 2.1, 1.5, and 1.17~MHz, the third, fifth, seventh, and ninth harmonics of the voltage signal coincide with the 10.5~MHz mechanical resonance, where the electro-optic response is enhanced by a factor of 4.2. The corresponding voltage spectra $U(f)$ are shown in Fig.~\ref{figAFC}(b) along with the corresponding adiabatic frequency shifts $\Delta\nu(t)$.

At a modulation frequency of 4.3~MHz, the measured frequency shift exhibits an almost perfect trapezoidal shape that repeats periodically. Only during the voltage ramp from 5~V back to 0~V, we do not observe adiabatic frequency tuning, as the resonator must first be emptied to avoid residual light from the preceding conversion cycle. Apart from this transition, the frequency shift increases linearly from 0 to 175~MHz and remains constant at this value for approximately 100~ns. This behavior is in excellent agreement with the expectation for a perfectly linear transfer from $U(t)$ according to Eq.~(\ref{eq:Linear}) with $n=2.13$ \cite{Gayer.2008} and $r=25$~pm/V \cite{Mendez99}. 

In contrast, at the other investigated modulation frequencies, the observed frequency shift deviates noticeably from the periodically repeated trapezoidal waveform. When the driving voltage is held constant at 5~V, the frequency shift does not remain constant but exhibits an additional modulation at 10.5~MHz, corresponding to the mechanical resonance frequency of the resonator. The amplitude of this modulation decreases with increasing order of the voltage-signal harmonic that coincides with the mechanical resonance. This behavior also nicely matches the prediction derived from the frequency-dependent electro-optic response discussed above.

These experimental results confirm that mechanical resonances can significantly affect the linearity of electro-optically driven adiabatic frequency conversion. When harmonics of the driving voltage coincide with mechanical modes of the resonator, additional oscillations appear in the frequency shift, even at higher harmonics that are more than two orders of magnitude smaller than the fundamental one. This behavior is consistent with the predicted frequency-dependent electro-optic response.

\section{Conclusion}

We have shown that mechanical resonances can strongly affect the linearity of electro-optically driven adiabatic frequency conversion. Pronounced deviations from linear behavior occur even when the amplitude of the higher-order harmonic of the voltage signal that coincides with a mechanical resonance is more than two orders of magnitude smaller than that of the fundamental component. Such resonances naturally arise in non-centrosymmetric materials that exhibit both the linear electro-optic and piezoelectric effects.

Linearity in the transfer from a periodic driving voltage $U(t)$ to a corresponding periodic frequency shift $\Delta\nu(t)$ can, however, be maintained under specific spectral conditions. If the mechanical resonances are sufficiently narrow, they can be placed between the harmonic components of the driving signal, where no significant spectral power is present. If the mechanical resonances are broader, they must instead be shifted to higher frequencies - well beyond the range of the relevant voltage harmonics. In this case, the amplitudes of the corresponding spectral components are intrinsically small, ideally by more than two orders of magnitude compared with the fundamental.

Although this study focuses on adiabatic frequency conversion, the conclusions are equally relevant for other electro-optically tuned frequency-control schemes. One example is self-injection locking of lasers to electro-optically controlled microresonators. For frequency-modulated continuous-wave (FMCW) LiDAR systems, periodic triangular voltage waveforms are used to generate corresponding optical frequency sweeps \cite{SnigirevUltrafast}. In all such cases, mechanical resonances that spectrally overlap with harmonics of the driving voltage can severely degrade the linearity of the frequency tuning process.

Our findings highlight the importance of considering mechanical resonances in the design of electro-optically tuned photonic systems and pave the way for more predictable and highly linear frequency control in resonator-based devices.


\begin{backmatter}
\bmsection{Funding}
We acknowledge support by the Open Access Publication Fund of the University of Freiburg. This work was financially supported by the German Research Foundation, DFG (Grant No. BR 4194/12-1) and Federal Ministry of Research, Technology and Space, BMFTR (Grant No. 13N16555).


\bmsection{Disclosures}
The authors declare no conflicts of interest.

\bmsection{Data Availability Statement}
The data that support the findings of this study are available from the corresponding author upon reasonable request.

\end{backmatter}


\bibliography{sample}

\end{document}